\documentclass{sigchi}

\toappear{\scriptsize Permission to make digital or hard copies of all or part of this work for personal or classroom use is granted without fee provided that copies are not made or distributed for profit or commercial advantage and that copies bear this notice and the full citation on the first page. Copyrights for components of this work owned by others than ACM must be honored. Abstracting with credit is permitted. To copy otherwise, or republish, to post on servers or to redistribute to lists, requires prior specific permission and/or a fee. Request permissions from permissions@acm.org. \\
{\emph{CHI 2016}}, May 7--12, 2016, San Jose, California, USA. \\
Copyright is held by the owner/author(s). Publication rights licensed to ACM. \\
ACM ISBN 978-1-4503-3362-7/16/05\ ...\$15.00.\\
http://dx.doi.org/10.1145/2858036.2858121}

\usepackage{balance}  
\usepackage{graphics} 
\usepackage{txfonts}
\usepackage{times}    
\usepackage[pdftex]{hyperref}
\usepackage{color}
\usepackage{textcomp}
\usepackage{booktabs}
\usepackage{ccicons}
\usepackage{todonotes}
\usepackage{tabularx}
\usepackage{arydshln}

\makeatletter
\def\url@leostyle{%
  \@ifundefined{selectfont}{\def\UrlFont{\sf}}{\def\UrlFont{\small\bf\ttfamily}}}
\makeatother
\def\pprw{8.5in}
\def\pprh{11in}

\definecolor{linkColor}{RGB}{6,125,233}
\hypersetup{%
  pdftitle={SIGCHI Conference Proceedings Format},
  pdfauthor={LaTeX},
  pdfkeywords={SIGCHI, proceedings, archival format},
  bookmarksnumbered,
  pdfstartview={FitH},
  colorlinks,
  citecolor=black,
  filecolor=black,
  linkcolor=black,
  urlcolor=linkColor,
  breaklinks=true,
}

\begin{document}

\title{Atelier: Repurposing Expert Crowdsourcing Tasks\\ as Micro-internships}

\numberofauthors{1}
\author{%
  \alignauthor{Ryo Suzuki$^1$, Niloufar Salehi$^2$, Michelle S. Lam$^2$,\\ Juan C. Marroquin$^2$, Michael S. Bernstein$^2$\\
    \affaddr{$^1$University of Colorado Boulder, $^2$Stanford University}\\
    \email{ryo.suzuki@colorado.edu, niloufar@cs.stanford.edu, \{mlam4, juanm95\} @stanford.edu, msb@cs.stanford.edu}}\\
}

\maketitle

\begin{abstract}
Expert crowdsourcing marketplaces have untapped potential to empower workers' career and skill development. Currently, many workers cannot afford to invest the time and sacrifice the earnings required to learn a new skill, and a lack of experience makes it difficult to get job offers even if they do.
In this paper, we seek to lower the threshold to skill development by repurposing existing tasks on the marketplace as mentored, paid, real-world work experiences, which we refer to as micro-internships.
We instantiate this idea in Atelier, a micro-internship platform that connects crowd interns with crowd mentors. Atelier guides mentor--intern pairs to break down expert crowdsourcing tasks into milestones, review intermediate output, and problem-solve together.
We conducted a field experiment comparing Atelier's mentorship model to a non-mentored alternative on a real-world programming crowdsourcing task, finding that Atelier helped interns maintain forward progress and absorb best practices.
\end{abstract}



\category{H.5.3.}{Group and Organization Interfaces}{}{}

%
%

\begin{figure}[t!]
\centering
  \includegraphics[width=1.0\columnwidth]{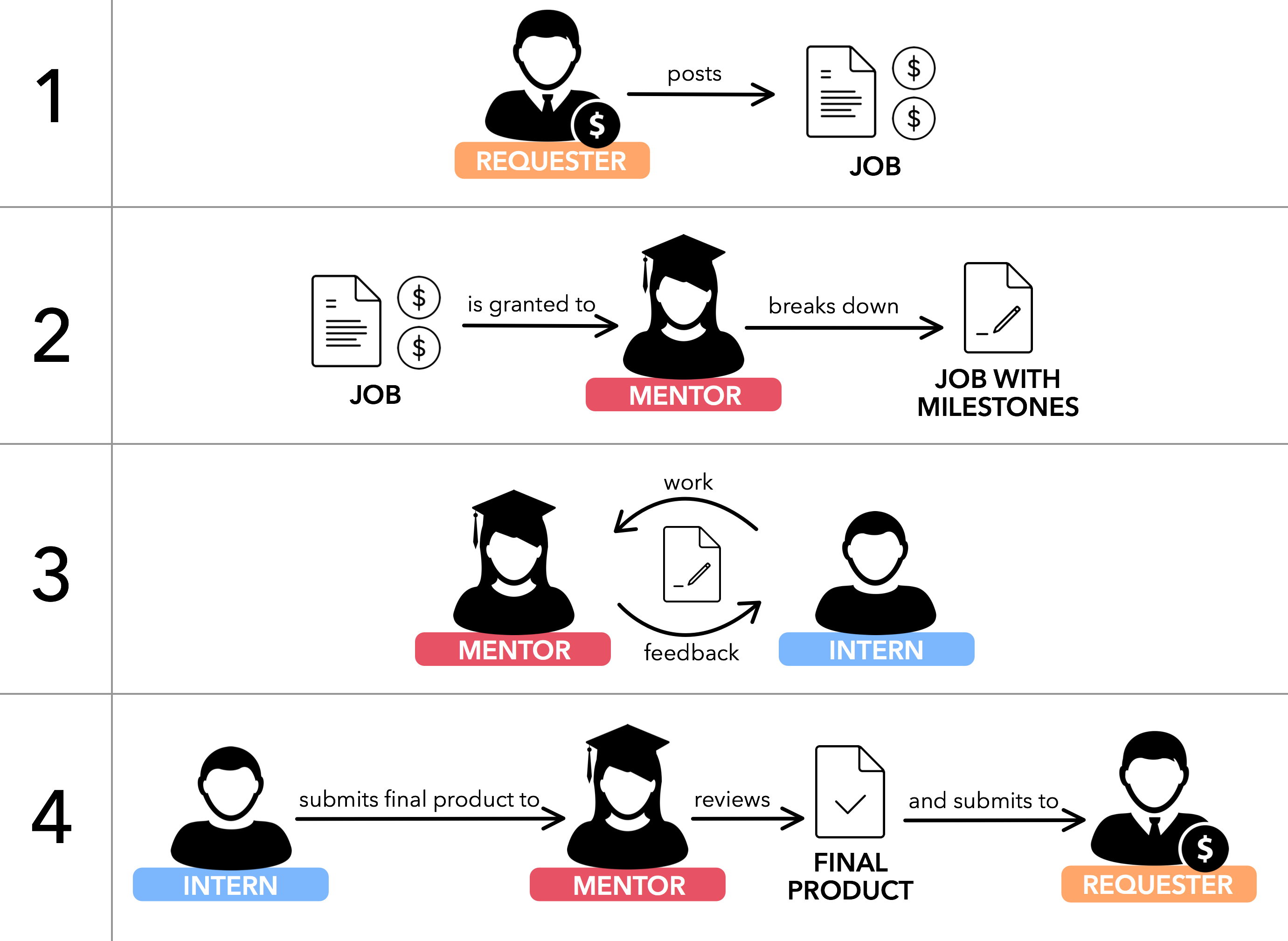}
  \caption{Atelier connects crowd workers (interns) with others on the crowdsourcing marketplace who have experience with a skill (mentors). It then facilitates a short micro-internship as the intern completes a real-world task from the marketplace under the tutelage of the mentor. (1)~Requesters post a task, then (2)~choose a mentor who breaks down the task into milestones. (3)~Intern works on the job while receiving feedback and guidance from the mentor, and finally (4)~submits the final product to the mentor, who reviews it and submits it to the requester.}~\label{fig:figure1}
\end{figure}
\section{Introduction}

For crowd work to stand as a viable long-term career option \cite{kittur2013future}, online crowd experts \cite{retelny2014expert} must be able to grow and continually refresh their skills. Traditional workplaces utilize on-the-job training and internships to enable employees' skill development while providing financial support. Crowd workers, however, are disincentivized from learning new skills: time spent learning is time spent \textit{not working}, which reduces income \cite{martin2014being}. Even if a worker does spend time learning new skills, platforms do not make it easy for the investment to pay off: it is difficult to get hired for new skills because expert crowdsourcing marketplaces have no ratings for the workers in their new skill areas \cite{kokkodis2015reputation}. As a result, many workers' skills remain static, and workers today often view crowdsourcing marketplaces as places to seek temporary jobs for their pre-existing skills rather than as venues for long-term career development \cite{kokkodis2015reputation}. With online work capable of expanding to many full-time jobs \cite{blinder2006offshoring}, it is critical to articulate a clear vision for how crowd work and career development will integrate.

As a first step to address the broader challenge of career development in online work, we focus on the problem of skill development. Skill development in crowd work encompasses the need to learn, improve, and develop new skills while competing in a paid crowd marketplace \cite{dontcheva2014combining, dow2012shepherding}.

In this paper, we explore whether pre-existing crowdsourcing tasks can serve as scaffolds to develop new skills through paid, mentored, real-world work experiences, which we term \textit{micro-internships}. We survey workers on a popular expert crowdsourcing marketplace, Upwork, to understand workers' perspectives on career and skill development. Time and financial constraints were a major barrier, suggesting that payment would be necessary for workers to be able to carve out time. We then sampled tasks from the marketplace to estimate whether they could serve as a ready source of paid micro-internships. We found that roughly one-quarter of tasks effectively isolate an expert skill: for example, vacation photo retouching (Photoshop skills), creating a web page for a small medical practice (web programming), and translating Turkish news articles (language skills). 

Based on these results, we propose a model wherein workers (\textit{interns}) connect with a more experienced crowd expert (\textit{mentor}) who can guide, provide feedback, and vouch for the final task quality. This expert maintains their usual payment rate by mentoring for less time than it would take them to complete the job, while also exercising their intrinsic motivation to teach and reinforce their own skills. The mentors break down the task into milestones, provide feedback, answer questions, and interface with the task requester, in order to ensure the quality of the final product. Interns learn by applying their skills to achieve a real-world goal \cite{papert1980mindstorms, schank1999learning}, get paid for their time, and gain a portfolio item and rating feedback to help break into the new area.

To demonstrate this concept, we present Atelier, a system to support micro-internships for crowd workers. Atelier sits complementary to the Upwork expert crowd marketplace. Experts apply to become mentors, and the requester chooses a mentor who they trust will ensure high-quality results. The mentor uses the platform to author intermediate milestones, and chooses an intern from a list of applicants. The mentor and intern agree on office hours, communicate via Atelier's synchronous chat, and use the system to review the work. When the intern submits the final work product, the mentor reviews it and returns it to the original requester for both payment and Upwork feedback rating for the intern.

In a field experiment comparing Atelier mentorship to a non-mentored experience, mentors successfully helped interns move forward when they got stuck, and introduced helpful resources and new technologies. Heavier use of Atelier for feedback was associated with higher quality work outcomes. Expert mentors improved the quality of work through sharing industry conventions and best practice. 

\section{Related Work}
In this section, we first review literature on career development in marketplaces for online labor. We then draw upon literature in education, with a focus on methods of learning through real-world tasks and mentorship, and adapt these methods to the online crowd work environment.

\subsection{Careers in expert crowd work}
The development of career ladders is vital to achieving a prosocial future for crowd work \cite{kittur2013future}, but currently long-term advancement in crowd work is difficult. One barrier to career development is reputation, since reputation is one of the foremost concerns of workers  \cite{martin2014being}. Since the availability of higher wage tasks depends on prior ratings, workers focus on maintaining a good reputation \cite{gupta2014turk}. This focus can discourage workers from attempting new tasks that extend beyond their comfort zone and could lower their ratings.

Even if workers successfully learn new skills and have the confidence to venture into a new class of jobs, it is still quite difficult for them to break into a new skill area of the market. This occurs because (1) reputation in prior fields of expertise cannot be transferred to a new field (e.g. from web development to graphic design) \cite{kokkodis2015reputation} and (2) requesters tend to rely on observable characteristics (e.g. work experience and certifications) to minimize risk in hiring decisions. Since workers' newly-developed skills are unobservable, they are overlooked \cite{kokkodis2014utility}. The primary motivation of most workers is monetary \cite{antin2012social, ipeirotis2010demographics, kaufmann2011worker, martin2014being}, and workers often sell their time because of limited opportunities for higher wages in the traditional job market \cite{martin2014being}. As a result, workers may be discouraged from learning new skills since subsequent employment in a new field may be uncertain.

Therefore, Atelier helps workers overcome the barrier to entry in new fields by enabling them to build their reputation and portfolio in new skill areas.
By furnishing workers with on-the-job experience, a high-quality portfolio item, and reputation relevant to their newly developed skillset, Atelier guides workers in their process of skill growth.

\subsection{Learning by doing}
Atelier's mentorship design is based on prior studies in education that indicate how learners benefit from both real world examples and the support of mentorship.

\subsubsection{Learning through real world tasks and mentorship}
The theory of situated cognition holds that conceptual understanding is inseparable from its applicable context \cite{brown1989situated}. While concepts can be acquired like tools, they are not actuated until they are used in a new situation --- concepts are continuously molded by the activities and communities in which they are used \cite{bransford1999people}. This educational notion points to the pedagogical method of cognitive apprenticeship, which focuses on the cognitive processes underlying the completion of complex real-world tasks \cite{collins1989cognitive}. In this method, apprentice learners take on authentic domain tasks and learn by observing and practicing with an expert mentor \cite{collins1989cognitive}. The cognitive apprenticeship approach has had great success in applications like high schoolers learning design skills \cite{liu1998study} and new Facebook engineers learning the company's code libraries \cite{bosworth2009facebook}.

Systems like LevelUp for Photoshop \cite{dontcheva2014combining} integrate real-world tasks into a volunteer learning process on the web. We extend this notion by sourcing tasks from a live marketplace and incorporating direct mentorship, which is often highly beneficial for high level tasks such as programming \cite{collins1989cognitive}. The most important elements of cognitive mentorships are that they (1) demonstrate the legitimacy of learners' intellectual understanding, (2) illustrate the failings of absolute heuristics and instead point to situational adaptation, and (3) engage learners as acting, culturally relevant members of a community \cite{brown1989situated}. Thus, Atelier not only presents learners with real industry tasks, but also implements their final products in the real world to situate the mentorship in a meaningful context. Atelier also encourages interns to complete multiple tasks so they can adapt their conceptual framework to various situations.

\subsubsection{Effective mentorship design}
The educational benefits of real world tasks are most evident when paired with an expert mentor, and one-on-one tutoring has been established as one of the most successful ways for learners to develop mastery \cite{ambrose2010learning, bloom19842, collins1989cognitive}. 
Our challenge is to provide the learning benefits of mentorship to a large pool of crowd workers while using expert mentors recruited from crowd work platforms and minimizing mentor time commitment. Prior work has shown that the introduction of tools to facilitate mentor--intern interactions can increase the availability and productivity of mentors in programming and creative fields \cite{guo2015codeopticon, rees2015building}. We therefore wanted to identify core traits of effective mentorship and foster these traits in Atelier. 

We focus on tasks in workers' zone of proximal development (ZPD), the class of tasks that extends beyond what an individual could achieve on their own, but could be made achievable through appropriate outside guidance \cite{vygotsky1978interaction}. This level of challenge is considered optimal for learners' benefit \cite{wood1976role}, and from this idea stems the concept of educational ``scaffolding'': the support, instruction, or resources that enable learners to achieve what they otherwise could not accomplish \cite{davis2004explorations}. The projects that interns encounter on Atelier, by nature, go beyond their existing skillset, but with the help of a mentor and a set of planned milestones, they become achievable 
\cite{wass2014sharpening}. 

Successful mentorship scaffolding consists of continuous diagnosis and readjustment of support, a variety of types of support, and gradual removal or ``fading'' of support to enable self-sustainability in future tasks \cite{stone1998metaphor}. Atelier builds upon these principles to maximize the effectiveness of a streamlined mentorship. A milestone breakdown of the project provides structure to the scaffold by keeping interns on track, keeping mentors updated on intern progress so they can diagnose intern issues, and providing modular steps that mentors can adjust according to an intern's specific needs \cite{wood1976role}. Various means of assistance like synchronous chat, question-answering, and video chat allow mentors to guide interns. Mentor feedback on the project, which has been shown to improve result quality, further adds to the variety of the scaffolded support \cite{dow2012shepherding, haas2015argonaut, kulkarni2012mobileworks, luther2015structuring}. Office hours (short sessions of directed mentorship) provide fast feedback that improves educational benefit \cite{kulkarni2015peerstudio}, but prevent over-dependence on the mentor and instead develop intern learning and autonomy. Atelier incorporates the core elements of effective mentorship scaffolding to increase the educational value of real-world tasks, develop applicable skills, and enable career growth for crowd workers.

\section{Surveying Workers' Career Opportunities}
We envision that crowd workers can potentially build their entire careers online \cite{kittur2013future}, but the literature knows little about whether workers already do so. Thus, we conducted a survey on the Upwork expert crowdsourcing marketplace to develop an understanding of workers' long-term needs and career goals. To recruit workers, we messaged 200 active workers across nine specialties (Graphic Design, Content Creation, User Interface/User Experience Design, Back-End Development, Front-End Development, Android Development, Photo/Video Editing, Video Transcription, and Quality Assurance Testing). We asked them to complete a 30-to-45 minute online survey for \$15, and we received 96 survey responses. (Male: 71, Female: 25; Age: 18--66, Median Age: 25) These workers were representative of active Upwork users who had a stake in the platform. Of these respondents,  the highest form of education received ranged from some high school (1\% of workers) to high school (9\%) to university (53\%) to graduate school (36\%); 8\% of workers had less than one month of Upwork experience, 40\% had one--to--twelve months of experience, and 50\% had more than one year of Upwork experience.

In the survey, we asked freelancers about the potential for growth in their Upwork careers, the ability to broaden their skillsets, and the training they would need to develop new skills. Participants provided binary responses regarding whether they felt they would stay on Upwork, whether they felt they could grow their career on Upwork, and whether they felt they could broaden their skillset. They also provided open-ended responses to explain their choices. To evaluate these responses, we engaged in an inductive process to iteratively read submissions, identify themes, code the responses with themes, and repeat to revise the themes. Three trends emerged from the survey results: (1) Workers want their careers to stay on Upwork, (2) Workers want to grow their careers on Upwork, but face job instability due to heavy competition, and (3) Workers want to broaden their skillset, but don't have the time or financial resources to do so.

\begin{table}[tb]
\newcolumntype{S}{>{\hsize=.15\hsize}X}
\newcolumntype{L}{>{\hsize=.85\hsize}X}
\newcommand{\mc}[2]{\multicolumn{#1}{|c|}{#2}}
\setlength\extrarowheight{5pt}
\begin{tabularx}{\columnwidth}{|L|S|}
\hline
\multicolumn{2}{|l|}{
{\bf Workers want their careers to stay on Upwork}
} \\
\hline
Workers whose careers were primarily based on Upwork or other crowdsourcing platforms & 61\% \\
Workers who wanted to stay on Upwork for the rest of their career & 67\% \\
\hline
\multicolumn{2}{|l|}{
{\bf Workers want to grow their careers on Upwork}
} \\
\hline
Workers who felt that they could grow their careers on Upwork & 89\% \\
Workers who described Upwork careers as short-term/temporary/unstable & 40\% \\
\hline
\multicolumn{2}{|l|}{
{\bf Workers want to broaden their skillset}
} \\
\hline 
Workers who wanted to broaden their skillsets on Upwork & 94\% \\
Workers who felt they \textit{couldn't} learn a new skill in the next 6 months & 39\% \\ 
\hline
\end{tabularx}
\caption{Our survey of 96 Upworkers indicated a strong affiliation to the platform as a long-term career option, despite difficulty learning skills.
\label{tab:survey}}
\end{table}

\subsection {Workers want their careers to stay on Upwork}
Freelance work constitutes a significant portion of these workers' careers: 
61\% of the surveyed workers relied almost entirely on Upwork or other online crowd work platforms for their livelihood; and 39\% depended primarily on a traditional job. Not only do workers heavily rely upon Upwork, but they also plan to stay on the platform: we found that 67\% of workers (42\% who were online freelancers plus 25\% who held a traditional job) wished to remain on Upwork for the rest of their career. Given workers' reliance upon online crowd work, we designed Atelier to augment the learning benefit of existing online crowd work tasks and target the development of practical skills that can be immediately and directly utilized for crowd work. Our results are consistent with prior survey results indicating that many freelancers would not quit freelancing to work with a traditional employer no matter the pay 
~\cite{upwork2015freelancing}.

\subsection {Workers want to grow their careers on Upwork, but face instability due to heavy competition}
Fully 89\% of workers felt that they could grow their careers on Upwork. This \textit{desire} for career growth, however, differed from workers' perceptions of their ability to actually \textit{achieve} career growth: many workers who wanted to stay on the platform voiced uncertainty about their ability to stay and succeed. In fact, the characterizations of Upwork careers were quite similar between those who did and did not want to stay on the platform --- 40\% of the surveyed workers described Upwork careers as short-term, unstable, or lacking in opportunity for growth. Workers said that they view it as  \textit{``a supplement, not a full time job,''} that they \textit{``hope to find something more regular,''} and that \textit{``there's nothing with trajectory into the field I really want to pursue.''}

When asked to describe barriers to career growth on Upwork, many workers also mentioned heavy competition for jobs (\textit{``I have to bid against more experienced, higher rated freelancers,'' ``Competing for jobs with over 10 applicants becomes discouraging,''}), and many workers mentioned an excess of underpaid jobs and low bidding prices (\textit{``[Clients set rates] far less than what the current minimum wage is,'' ``I cannot compete with low bids set by [other] freelancers,'' ``It is horrifying that some clients think that they are entitled to get a considerable amount of work done for such small rates''}). These obstacles prevent workers from growing their online freelance work from part-time financial supplements into long-term, viable careers. Thus, we designed Atelier to build career trajectories and allow workers to transition between skill domains in such a way that they aren't in direct competition with experts in that domain.

\subsection {Workers want to broaden their skillset, but don't have time or financial resources to do so }
An overwhelming portion of workers expressed a desire to develop and expand their skills --- 94\% of workers wanted to broaden their skillset on Upwork, but many of these workers felt that they could \textit{not} learn a new skill within the next six months; many of them didn't know how to begin learning such a skill. The most commonly cited reason was that workers felt they did not have the time or financial resources to learn a new skill, especially because they lacked the ability to sacrifice working hours for unpaid learning hours (\textit{``Because I need more money to support my family, I take all the time I have for work. This gives no room for me to have time to enroll myself in a class where I can improve my skills,'' ``If I'm [learning], I'm not working on freelance jobs that bring in some money''}). Additionally, once a worker develops new skills, it is difficult to break into the new market: 77\% of the workers found it difficult to convince clients to give them their first few contracts when they first joined Upwork. Workers attributed this difficulty to their lack of ratings, logged hours, prior feedback, or portfolio material upon joining the site. To address these needs, Atelier was designed to establish reputation, build portfolios, and provide compensation to make learning viable for these workers.

Overall, workers generally felt that Upwork did not adequately aid them in development of new skills. When asked what kind of feedback or training would be most helpful to develop new skills, it was surprising that many workers requested online courses. Despite the multitude of online resources and courses available today, the workers still felt that they needed more resources to expand their skillset --- or perhaps those resources felt too distant from the lived realities of Upwork work. In addition, many workers requested one-on-one training or mentorship involving personal feedback, and they also requested real-world jobs similar to those they aspire to work on. We therefore focused our model on a unification of the latter two methods --- expert mentorship and real-world projects --- to provide manageable, concrete steps for workers, who often don't know how to get started on skill development. We then turned to Upwork to evaluate the suitability of existing jobs for this purpose. 

\begin{figure*}[t!]
  \centering
  \includegraphics[width=2.1\columnwidth]{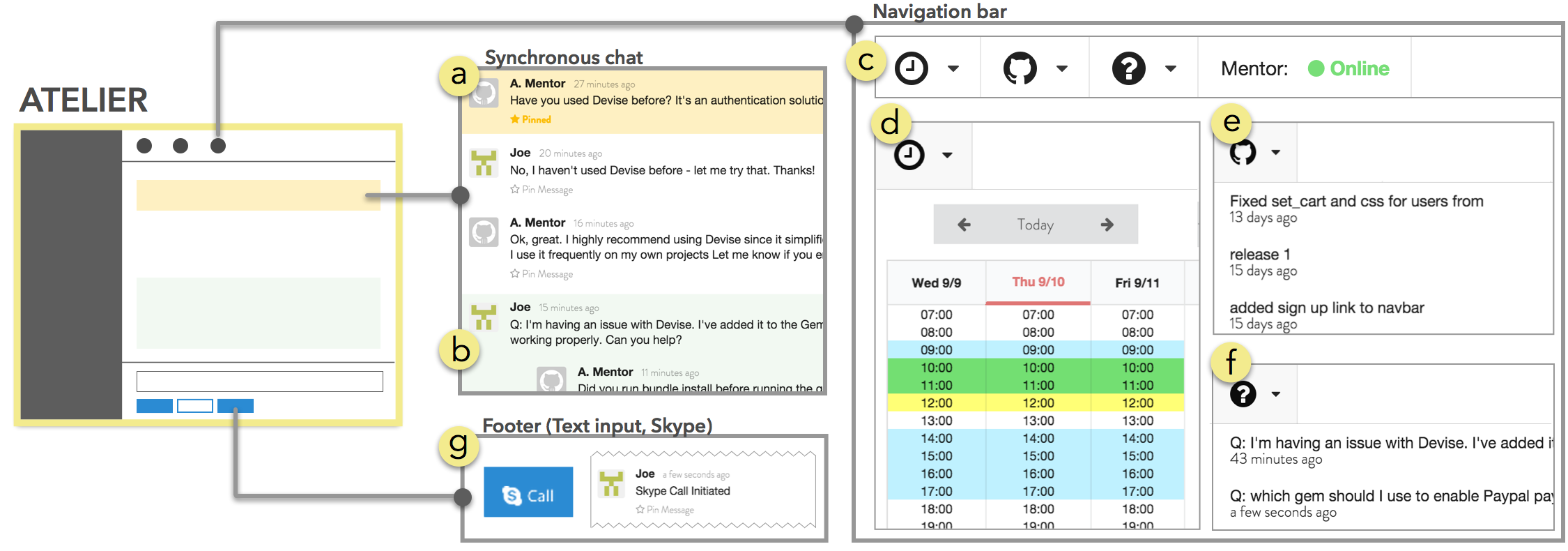}
  \caption{Atelier enables mentorship through (a)~synchronous chat between interns and mentors, (b)~question highlighting and threading, (c)~a notifications bar that hosts (d)~office hours, (e)~Github commits, (f)~unanswered question notifications, and (g) video calls.}~\label{fig:figure2}
\end{figure*}

\section {Analysis of Learning Opportunities on Upwork}
With categories as diverse as Architecture, Design, Marketing, and Software Development, Upwork hosts myriad task postings. But could they serve as effective internships, by isolating skills within a bounded time window? To see if there are sufficient tasks that would make a platform like Atelier feasible, we conducted a feasibility study of learning opportunities on Upwork. If even a small minority of 5--10\% of tasks could be supported, there would be an opportunity to leverage the marketplace for skill development. We found that the most effective tasks to meet worker's learning needs were those that were geared toward entry- to intermediate-level workers and whose projected length fell in the scope of hours, days, or weeks. Furthermore, tasks that best fit Atelier and its goals tend to have (1) specific overarching goals, (2) deliverable end-products, (3) non-urgent deadlines for completion.

\begin{table}[b]
\newcolumntype{S}{>{\hsize=.4\hsize}X}
\newcolumntype{M}{>{\hsize=.6\hsize}X}
\setlength\extrarowheight{5pt}
\begin{tabularx}{\columnwidth}{|M|S|}
\hline
{\bf Criteria for Micro-internship} & {\bf Proportion}\\
\hline 
Specific Goals & 86\% (105/120) \\
Deliverables & 83\% (99/120) \\ 
Non-urgent Deadlines & 71\% (85/120) \\
Not trivial or invalid & 78\% (94/120) \\
\hdashline
All Criteria  & 42\% (50/120) \\ 
\hline
\end{tabularx}
\caption{Tasks that meet the criteria to be a potential micro-internship. \label{tab:analysis}}
\end{table}

To determine the existence of a significant pool of Atelier-suitable task postings, we took a random sample of Upwork job postings in summer 2015. First, we limited the search to tasks listed at an Experience Level of \textit{Entry} to \textit{Intermediate} and Project Length of \textit{Hours} to \textit{Weeks}. These queries resulted in a total of 44,258 tasks (out of 73,879). We collected a sample of 120 of these tasks by randomly selecting ten from each of the twelve Upwork categories, including web \& software development; writing; sales \& marketing; and admin Support.

We then hired a worker from Upwork who evaluated the tasks in our sample across three criteria: (1) specific goals, (2) deliverables, and (3) non-urgent deadlines. The expert also noted tasks that were trivial, meaning the task did not require a mastered skill to perform, and those that were invalid, meaning the job did not have enough information to decipher what task was being advertised. A task was considered satisfactory for Atelier if it satisfied all three criteria and was neither trivial nor invalid. We computed the inter-coder reliability (Cohen's Kappa) between a member of our research team and this expert across 24 jobs from our sample. We calculated a Kappa value for each criterion: specific goals ($\kappa$=0.6), deliverable ($\kappa$=0.7), non-urgent ($\kappa$=0.5), and trivial/invalid ($\kappa$=0.8). These values represent a fair to good agreement between the expert and our team.

This process resulted in 42\% of sampled tasks (50/120) meeting the criteria to be a micro-internship (Table~\ref{tab:analysis}). Extrapolating to all tasks on Upwork, specifically those excluded from the sample due to experience level and project length requirements, results in an estimated 25\% (0.25 = 0.42 * (44,258 / 73,879) ) of Upwork tasks as potential micro-internships.

\section{Atelier}
Atelier sources learning tasks and mentors from crowd work platforms to help workers develop new skills. Atelier, an AngularJS and Ruby on Rails web platform, was designed through an iterative process: we recruited pairs of mentors (topic experts with high ratings) and interns (novice to the topic area) from Upwork, then asked them to collaborate through successively refined prototypes as we observed, redesigned, and iterated on the system.

\subsection{Connecting mentors and interns with existing tasks}
Atelier allows requesters to cross-post their Upwork jobs to the platform in order to make them available as  micro-internships. Requesters --- those who are willing to pay to get work done --- post their tasks to Atelier, indicating the types and levels of expertise they require, as well as the skills that they hope the intern might learn. Experts then apply to mentor the position by submitting their Upwork profile as well as a short application note describing their qualifications. After several mentors apply, the requester chooses the mentor who they wish to oversee the task. 

Much like traditional offline internships, there are several reasons a requester or mentor may wish to get involved. Requesters may be motivated to save some money for hiring less experienced workers, while still having assurances that an expert will review the final product. They may also be interested in identifying future superstar workers early so that they can retain them. Finally, they may view micro-internships as a pro-social act toward the crowd work ecosystem, wherein they help train and support the workforce. From the mentors' perspective, our users reported that they found it a good source of extra income that did not require a heavy time commitment. In practice, they also often carried intrinsic motivations to mentor others and share their knowledge.

Expert mentors' hourly rates are typically three to four times interns' hourly rates, so Atelier must divide up the payment to balance their competing financial needs. Our purpose in this research is not to develop a business model (much less a \textit{good} business model), but it is important to establish a feasible proof-of-concept that can scale and stand on its own. So, by default, Atelier gives half of the task payment to the mentor in exchange for hours of mentorship, and the other half goes to the worker for completing the task. For example, for a \$300 web development task, experienced web developers (\$30-\$50 per hour) are paid \$150 for a few hours of their mentorship time distributed over the project, and interns (often \$10 per hour) are paid \$150 for their part in completing the task. Therefore, both parties make their typical going rate --- and we aim for the interns to make \textit{more} money than they would have in their previous skills, which are typically lower-paying than the skill they are trying to learn. 

Interns browse Atelier to find jobs to help solidify their skills. The expectation is that they have taken any relevant training they need to be prepared (e.g., a Codecademy course, a Massive Open Online Course). As we found in the Upwork survey, workers face heavy competition, especially when entering a new field, so Atelier separates these beginners into a class of \textit{interns}, so that beginners don't have to compete with domain experts upon learning a new skill. Interns apply to the position on Atelier with their Upwork profile and basic qualifications --- the intention is not that they be the most experienced worker on Upwork, but instead that they be dedicated to the task and have the necessary prerequisites. Mentors choose the intern they think is the best fit for the internship, set a deadline, and begin.

\subsection{Milestones: guiding interns, aiding mentor awareness}
Without scaffolding, interns took the wrong path to completing the task or got stuck and didn't know what to do. Worse, they failed to update mentors of their status \cite{rees2015building}, making it difficult for mentors to help. Learning succeeds best with concrete achievable goals \cite{locke1980goal}. We thus leverage the mentors' experience to break down each task into a series of milestones for the worker to follow and check off as they complete. For example, a Ruby on Rails web development task's milestones could be: \textit{1) Prepare project environment, 2) Static pages, 3) Gallery page, 4) User profile}. Mentors log into Atelier and transform the requester's task description into a set of milestones and steps within each milestone. For example, steps for the  milestone above are \textit{1-1)~Generate the initial website skeleton, 1-2)~Append your project to a git repository, 1-3)~Include a Postgres or MySQL gem to your project}. 

\begin{figure}
\centering
  \includegraphics[width=1\columnwidth]{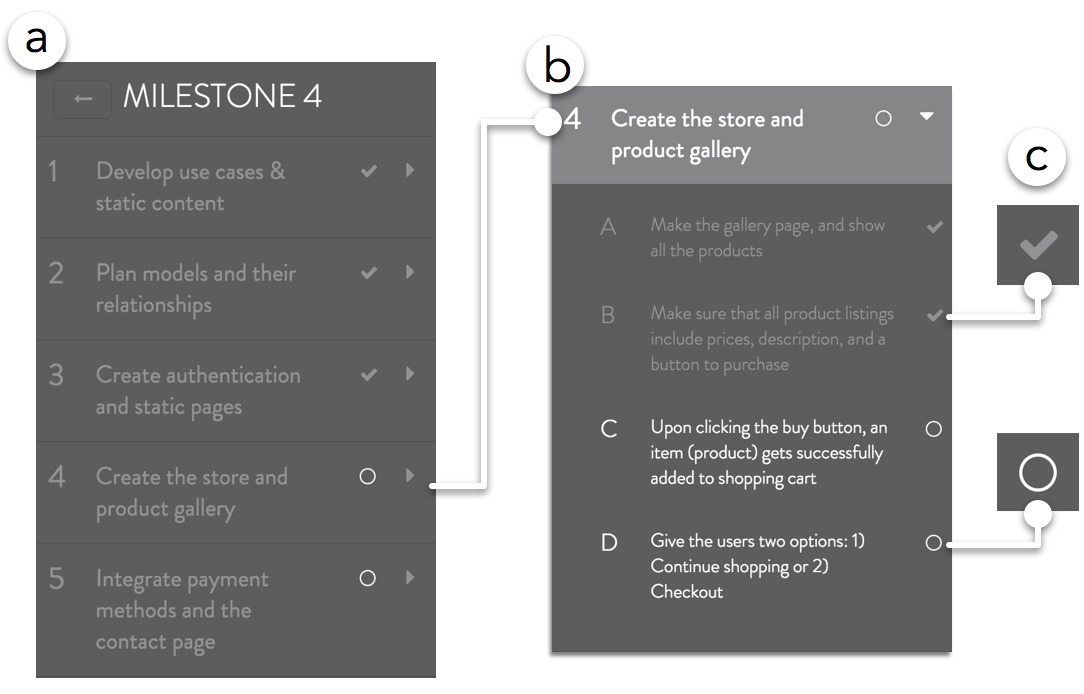}
  \caption{Mentors break down tasks into goals. (a)~Milestones are macro-scale guides. (b)~Steps are implementation-focused objectives. (c)~Interns can check off milestones and steps as they complete them.}~\label{fig:figure3}
\end{figure}

Milestones are macro-scale guides for completing the project and steps are more implementation focused. As an intern makes progress, she checks off each step or milestone in the Atelier interface (Figure~\ref{fig:figure3}). This ensures that the mentor is aware of the current state and progress. Reciprocally, if a long time has passed without the intern checking off a new step or milestone the mentor can check in to help unstick the intern.

Project-specific integrations may also be added to Atelier to further support collaborator awareness on progress. Currently, Atelier includes GitHub integration. This integration adds an update and link to the Atelier chat whenever a new commit is pushed to the project repository (Figure~\ref{fig:figure2}(e)). Other integrations are possible for different task domains (e.g., Behance for art, SoundCloud for audio, Google Drive for documents and files).

\subsection{Office hours: coordinating distributed workers}
Once milestones are established, the intern may begin the task with the support of the mentor. The mentor and intern share a chat channel for the project (Figure~\ref{fig:figure2}(a)). If text chat becomes difficult to use (e.g., for deictic references to code or art), the intern and mentor can utilize Atelier's Skype integration to launch a videocall and screen share (Figure~\ref{fig:figure2}(g)).

It is well known that virtual teamwork suffers from coordination challenges \cite{hinds2002distributed, hollan1992beyond}. 
Without active mentor intervention, interns proceed without asking for help and can make poor decisions as a result \cite{rees2015building}. Mentors are also often more sensitive to deadlines than interns, since mentors are more acculturated to Upwork's norm of deadline adherence. 

To balance mentors' availability with interns' efforts, we considered three possible options: an always-on chatroom, coaching sessions, and office hours. 
An always-on chatroom enables the intern to ask for help whenever the mentor drops in, but this approach does not guarantee interns that the mentor will be available for synchronous help, and in practice, leads mentors to dedicate many more hours than their stipend supports. Another option is coaching sessions, where the intern shares their screen while they work and the mentor observes and interjects to help unstick the intern when necessary. This approach gives the mentor more visibility into the intern's progress, but interns are still afraid to ask for help (or, potentially worse, they may assume they are on the right track if the mentor does not speak up). A third option is scheduled office hours, where both mentor and intern are expected to be available. Atelier pursues this third option, which sets a norm of direct coaching \cite{rees2015building}. To succeed at an office hours approach, Atelier must help mentor--intern pairs coordinate.

Atelier helps users establish office hours to alleviate the timezone problem, support direct coaching, and ensure that mentors are not overtaxed. To ensure that office hours are clear and fully utilized, they are scheduled immediately at the outset. Upon accepting the mentorship position, the mentor fills out their availability. When the intern joins the job, they share their availability. Overlapping mentor and intern times are set as office hours, and either user can quickly make subsequent adjustments that are visible to both users (Figure~\ref{fig:figure2}(d)). 
\begin{figure}
\centering
  \includegraphics[width=1.0\columnwidth]{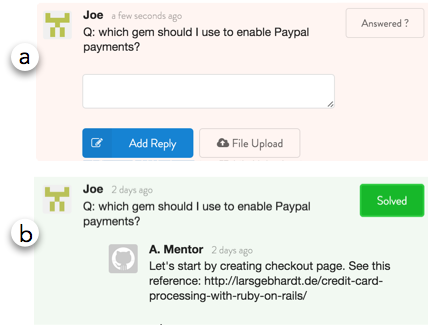}
  \caption{Interns can ask threaded questions within the chat. New questions produce notifications (See Figure~\ref{fig:figure2}(f)), (a) are highlighted in the interface, and (b) allow threaded replies.}~\label{fig:figure4}
\end{figure}
\subsection{Questions: getting unstuck}
Many interns mentioned that they found it difficult to clearly explain their issues when they lacked the necessary vocabulary or experience level \cite{furnas1987vocabulary}. Such questions are core to the learning process --- they are teachable moments \cite{schwartz1998time} at which the mentor can step in to smooth over conceptual misunderstandings or exemplify industry best practices and terminology. For example, one intern mentioned that he could learn how an HTTP request works through fixing bugs. However, these questions can easily become lost and overlooked in the midst of a fast-paced synchronous chat log. 

Atelier thus promotes questions as primary entities within the interface (Figure~\ref{fig:figure4}, Figure~\ref{fig:figure2}(b)). Rather than typing into the chat window as usual, an intern can submit a question. While the question remains unsolved, the question is visible via a notification badge at the top of Atelier and the mentor can click on the badge to jump to each unsolved question within the chat (Figure~\ref{fig:figure2}(f)). Unlike the rest of the chat, questions are threaded with replies to ensure that the question is directly addressed and the answers are found immediately below the original question. By distinguishing solved and unsolved questions and enabling direct replies, unsolved questions become actionable and urgent, and the intern's issues can be pinpointed and solved much more swiftly.

\subsection{Submitting: feedback and ratings}
Mentors typically take on the role of managing the project to achieve the deadline. They also act as gatekeepers, returning the work to the intern for revision as necessary until it is up to the requester's standards. When the mentor judges the work to be complete, they submit it to the requester for feedback. The requester then provides constructive feedback to the mentor and intern, which the mentor helps interpret for the intern --- and when necessary, negotiates with the requester. This process iterates until the task is accepted.

At this point, the requester gives public feedback to the mentor and intern. The requester communicates with the mentor to decide on a rating for the intern, which they enter into Upwork and Atelier. Likewise, the intern and requester communicate about the rating to give the mentor. These two pieces of feedback help boost interns' and mentors' reputations on Upwork so that they can receive future work.
\subsection{Publishing: sharing the experience}
Atelier allows the mentor and intern to publish their chat transcript (anonymously) on Atelier for the benefit of others. These transcripts augment the original task with a successful breakdown of the project, expert feedback, and questions and issues addressed by an expert placed in the context of a project's history (from Github). 
This enhanced tutorial offers unique opportunities for novices readers to shadow expert workflows \cite{lafreniere2013understanding} and gain tacit knowledge which experts explicate in the process of guiding their interns \cite{nash2006tacit}. 

Mentors can also benefit from this publishing feature. Mentors can use published transcripts as a reference for future projects to reduce their teaching cost, being able to anticipate errors and get inspiration for potential milestone breakdowns. In addition, the tutorial allows the reader to pay the original mentor for help and feedback without leaving the page --- another reason why experts would want to mentor on Atelier.

\section{Evaluation}
To evaluate the effectiveness of Atelier's mentorship model, we conducted a controlled experiment comparing mentor-intern pairs to interns working alone. 

\subsection{Method}
As a representative task, we identified a web development job posted on Upwork: \textit{Create a Ruby on Rails web application for a company's e-commerce store}. 
This task fits the criteria in the feasibility study (specific goals, deliverables, non-urgent deadlines, and non-trivial). Furthermore, workers in the web development domain commonly attain much of their knowledge from online resources and this sort of background is what we envisioned as the proper entry point for Atelier. Experts and potential interns estimated the task at roughly twenty hours of work. The task had the following basic requirements: 1) a store and a gallery to show products with pictures and prices, 2) shopping cart function that can save products, 3) checkout function that allows payment via Paypal, Google Wallet, and major credit cards, and 4) contact us page that allows users to post comments and feedback.

We recruited 5 mentors and 22 interns from Upwork. We conducted a brief pre-survey to ensure that interns had taken an appropriate MOOC or had prior knowledge and experience of the assigned task with a limit of no more than two years of experience. Mentors were expert Upworkers who had extensive experience and high ratings. All interns were recruited using a single Upwork job post that sought Ruby on Rails novices to participate in a micro-internship. Interns were then randomized into either a \textit{control} condition, where they worked alone, or a \textit{mentorship} condition, where they were matched with one of the five mentors and used Atelier. To replicate realistic financial conditions, interns and mentors in the mentorship condition split the task earnings; interns in the control condition kept the entire amount (\$300).

Interns had ten days to complete the task. During the study, we observed mentor and intern interactions. 
To gather further data, we also surveyed our participants six days into the experiment and again at the end of the study. The surveys asked about overall impressions of Atelier, differences between the intern and mentor experiences, and changes in intern-mentor interactions in response to rising deadline pressures.

Finally, we recruited one additional Upwork Ruby on Rails expert responsible for evaluating all submitted E-Commerce projects.
The expert was blind to condition, and gave each project a rubric-based score out of ten, and ranked all fourteen results (1: worst rank, 14: best rank) based on the overall quality.
Using an external evaluator avoid biases where an individual involved in a product's development (e.g., the mentor) tends to have a positive bias in their evaluation of the product \cite{norton2011ikea}. However, to cross-check the validity of the expert's ratings, we also asked the mentors to give scores (5-point Likert scale) on their own project's outcomes.


\subsection{Participants}
All twenty-two interns were male (reflecting a sample bias in Upwork's programming categories) and 71\% of the interns either had a college education level or were in college; the remaining interns declined to respond their education level. Mentors had an average of four years of experience and a 4.7/5 rating on Upwork. Of the twenty-two interns originally hired, fifteen participants (eight mentored; seven non-mentored) submitted their final projects. After receiving the submissions, we discovered through the git log that one participant (non-mentored) completed his task by quietly recruiting friends to help, so we excluded this result. We therefore used fourteen valid participants (eight mentored; six non-mentored) for our evaluation. 

\subsection{Results}
Overall, although the score of the quality in the mentored condition was slightly higher (Median: 6.0/10, 1st Quartile (Q1): 5.0, 3rd Quartile (Q3): 6.3) than the non-mentored condition (Median: 5.5/10, Q1: 5.0, Q3: 6.0), due to large individual differences, the difference was not statistically significant (scores: t=0.22, p$>$0.05, ranks: t=0.25, p$>$0.05). Mentors had fair agreement in their assessment of project quality with the external rater ($\kappa$=0.24, p$<$0.05), though the mentors had a positive bias. 

As we describe in the next session, the core measurable impact was not due to the mere availability of Atelier, but with how interns and mentors utilized it. Additionally, the post-mentorship survey and our observations show that mentors helped the interns move forward when they got stuck and improved the quality of work by sharing industry conventions and best practices.

The interns in the mentored condition completed their tasks in 40.8 hours on average (SD=21.0) and the mentors' actual work time was 5.3 hours on average (SD=1.8). The task completion time in the mentored condition was slightly longer than the non-mentored condition (Mean: 37.4 hours, SD=21.6), but there was also no significant difference in this measure (t=0.28, p$>$0.7). Thus, mentors did achieve their advertised hourly rate by mentoring, though workers underestimated how long the task would take and thus underperformed their hourly rate in both conditions.

%


\begin{table}[t]
\newcolumntype{S}{>{\hsize=.3\hsize}X}
\newcolumntype{M}{>{\hsize=.4\hsize}X}
\setlength\extrarowheight{5pt}
\begin{tabularx}{\columnwidth}{|S|M S|}
\hline
{\bf Feature} & {\bf Mean} & {\bf SD} \\
\hline
Messages & 65.6 messages & (64.8)\\
\hline
Questions & 8.5 questions & (6.0) \\ 
\hline
Milestones & 6.5 milestones & (1.8) \\
\hline
Steps & 22.1 steps & (19.6) \\
\hline 
Office Hours & 6.6 sessions & (0.7) \\
\hline
\end{tabularx}
\caption{Use of Atelier's features. \label{tab:feature}}
\end{table}

\subsubsection{Heavier use of Atelier was associated with higher quality}
Though all mentor--intern pairs were provided with the Atelier system and were given detailed tutorials on how to use it, not all pairs made full use of Atelier's features (Table~\ref{tab:feature}). The extent to which mentor--intern pairs used Atelier for feedback purposes was associated with final product quality.

Questions and feedback were most directly related to educational scaffolding and mentorship support, whereas general messages sometimes were a bit less focused (e.g. introductions, greetings, goodbyes). 
Thus, we looked into the relative frequency of questions and feedback items since they were more indicative of constructive use of the Atelier platform than the total number of messages which varied between each mentor--intern pair. The quality of the main project (the final rank score, 1: worst rank, 14: best rank) had a strong positive correlation with the proportion of all messages that were questions ($\beta$=39.4 , $corr(x, y)$=0.84, t=3.84, p$<$0.01); the quality of the main project also had a strong positive correlation with the proportion of all messages that were in the broader category of either questions or feedback ($\beta$=42.7, $corr(x, y)$=0.89, t=4.68, p$<$0.01).

\subsubsection{Mentors helped unstick interns, provided conceptual guidance}
While working on the project, some interns got stuck due to blocking technical or conceptual issues, and they asked their mentor for help. Even though mentors were only available during Office Hours, they were able to unstick interns promptly. For Atelier's questions (highlighted, threaded questions), the median response time, amount of elapsed time between the original question and its first threaded response, was two hours (Q1: 6 minutes, Q3: 12.5 hours). Most of these questions were presented as technical issues:
\begin{quote}
Intern 2: \textit{Error: Showing \textless ERROR\textgreater  First argument in form cannot contain nil or be empty} \\
Mentor 1: \textit{add ``@comments = Comment.last(10)'' to new method} \\
Intern 2: \textit{...it works! thanks!}
\end{quote}

Although these technical problems may have been due to small errors, they were sometimes indicators of gaps in an intern's conceptual understanding. Mentors often provided more than just the solution itself --- they explained \textit{why} the issue had occurred and/or \textit{how} the solution worked. This guidance helped clarify interns' conceptual understanding and saved them a significant amount of time.  
\begin{quote}
Intern 4: \textit{I've learned how to communicate with an API [...] after my mentor figured out I was missing a ``?'' sign in the request [...] I searched for what I was doing wrong for hours regarding the ? and he spotted it in a few minutes [...] he said ``it was a basic HTML request issue'' and one I wouldn't probably figure out.}
\end{quote}
\begin{quote}
Intern 6: \textit{``I got great advice on how to think in an OOP way, and I was able to ask someone for advice when I couldn't find any searching the internet.''}
\end{quote}

Interns also felt they could generalize this new knowledge: \textit{``I've learned how to debug better''} [Intern 4], \textit{``I got more experience searching for problems solutions''} [Intern 6]. 

\subsubsection{Mentors suggested helpful resources and new technologies}
Mentors proactively suggested resources and technologies that they felt would be helpful for interns, and explained why these technologies were useful to this project.
\begin{quote}
Mentor 1: \textit{For static pages you can use High Voltage gem; it is light weight, well documented, and easy to use. [...] Have you heard about HAML (http://haml.info)? It is an easy markup language.}
\end{quote}

Furthermore, interns accepted these suggestions and used the resources recommended by their mentors, which helped them to succeed. For example, this intern was able to complete a full milestone by using resources suggested by his mentor:
\begin{quote}
Mentor 1: \textit{There are a lot of good tutorials how to create a cart model in Rails, e.g: [URLs of web tutorials]}\\
Intern 2: \textit{Nice tutorial. I finished Milestone 6.}
\end{quote}

When explaining why they felt their internship was a positive experience, interns explicitly mentioned the helpfulness of technologies suggested by their mentor (\textit{``My mentor introduced me to a lovely templating language --- Slim --- that I've now used and will keep on using''} [Intern 4]).

\subsubsection{Mentors improved the quality through sharing best practices}
Interns on Atelier received guidance from their mentor on industry conventions and best practices. Mentors often introduced these standards when giving feedback on the project or the intern's stated plan of action: 
\begin{quote}
Mentor 3: \textit{...make a seed file as it is more appropriate and according to Rails conventions. [...] write all the steps to start the project in your Github repository ReadMe; I know [it is] basic, but it's a good practice.}
\end{quote} 
\begin{quote}
Mentor 1: \textit{...in 99\% of cases, companies want show their HQ, so I suggest to move the HQ photo from the background and place it near the text.} 
\end{quote}
Interns who elaborated on reasons for their positive mentorship experience mentioned that mentor code feedback was helpful (\textit{``...learnt a lot of approaches regarding code implementation [from] my mentor''} [Intern 7]).

Sharing these industry conventions and best practices could have a positive effect on final project quality. For example, Mentor 1 and Mentor 4 recommended that their interns use the Braintree Rails gem for checkout functionality and active\_merchant gem for admin views. The evaluator gave these interns higher scores (7/10 and 8/10 respectively) based on these implementations. The evaluator noted: \textit{``Used Braintree for checkout, which works well and doesn't have any redirects. Checkout is simple and all on one page.''}; \textit{``Uses active\_merchant as the admin interface as well, allowing for easy user and product management.''}.

\subsubsection{Mentor behavior/quality greatly affected the intern experience}
All interns in the experimental condition received a mentor who had been deemed an expert, but not all mentors provided the same amount of support for their intern. The number of milestones and steps created by the mentor was one indicator of the amount of guidance that a mentor provided, especially since milestones comprise the very core of Atelier's scaffold. We log-transformed the number of milestones due to a skewed distribution, and found that the the quality of the main project (measured by the ranking score, 1: worst rank, 14: best rank) had a strong positive correlation with the log of the total number of milestones and steps provided to the intern ($\beta$=4.087, $corr(x, y)$=0.71, t=2.44, p$<$0.05). In other words, the more concrete the project breakdown, the higher quality the final project.

We also found that mentors spent more time to help low-ability interns. There was a significant negative correlation between the time mentors devoted and the rank of main project quality ($\beta$=-2.1397, $corr(x, y)$=-0.86, t=-3.701, p$<$0.02). Since interns only had a limited amount of mentor interaction, the main project quality for this single mentorship experience was more indicative of original skill level for low-ability interns; however, mentors were willing to devote more time to guide these less-experienced interns.

\subsubsection{Mentors exercised intrinsic motivations}
In our post-mentorship survey, all 5 mentors said that their role as a mentor was a positive experience, and 4 of the 5 mentors said that they felt their own skills were reinforced by the mentorship. In addition, their responses suggested that Atelier was beneficial and engaging because it allowed expert workers to share their expertise and gain teaching experience, e.g., \textit{``It was a good experience because I share my experience and knowledge with the interns''} (Mentor 4).
We asked mentors to indicate to what extent they were motivated by teaching their intern using a 7-point Likert scale (1: Not motivated at all, 7: Very motivated), and mentors responded that they were highly motivated (average: 6.8). Additionally, the average amount of time mentors spent working with each intern was 5.3 hours (SD = 1.8), and 7 of 8 mentors stated that the time commitment was reasonable. 


\section{Discussion and Future Work}
Our evaluation of Atelier suggests that it can connect experienced mentors to novice interns: the system enabled mentors to successfully unstick interns, teach them best practices, and guide them to successful completion of the task. In this section, we reflect on limitations in our evaluation as well as opportunities for future work.

\subsection{Feasibility}
Although developing a business model is not central to this work, it is also important to discuss the real-world feasibility of Atelier achieving a stable equilibrium. Interns gain financial and educational benefits from Atelier; mentors' incentives are not as clear. Would mentors be willing to participate and thus share payment with an intern? Would mentors be discouraged because of a potential increase in competition and decrease in jobs?

Our main insight here is that the time commitment for highly skilled mentors on Atelier is significantly less than would have been required for them to complete the task themselves; making mentorship on Atelier economically feasible. In our study, mentors reported spending 5.3 hours on average on the task, which translates to roughly \$30 per hour. This rate is consistent with these experts' advertised hourly rates (average: \$29.4/h) and all mentors responded that the time commitment was reasonable. In contrast, experts estimated that it would take them 20 hours to complete the task themselves which would result in {\raise.17ex\hbox{$\scriptstyle\sim$}}25\% of the mentor's desired wage, and they would not accept the job. On the other hand, interns are still ``on their way up'' to commanding higher wages and could not have completed the task without the mentors coaching. Therefore the division of wages benefits both. 
Additionally,  Atelier's model can potentially produce more work than current markets for experts, because it makes many currently-infeasible (too low-priced) jobs available to them as a mentoring opportunity.

One challenge with Atelier: if a project fails, who is responsible, and how does the requester or the platform hold them accountable? One user in the mentored condition dropped out before starting the project, and one participant dropped out midway through the project.  These failures may have occurred due to the lack of time management. The ability to follow through with responsibilities depended on whether interns were able to manage their time, but mentors had little control over this factor. 
One way to manage the progress is to leverage the milestones and steps feature and enable the mentor to set more detailed deadlines as fine-grained checkpoints. 

Another method to regulate responsibility might be to allow mentors and interns to rate each other. However, poorly designed rating systems may be problematic. For example, in our deployment one intern claimed that his mentor was ``mostly non-existent'', despite the fact that his mentor had among the highest rates of message exchange on the platform. This statement may have arisen from frustrations about a short period close to the project deadline during which the mentor was unreachable. This problem may be addressable via a per milestone rating system that would allow interns and mentors to provide continual feedback. Upwork and other online crowd work platforms face similar challenges in their rating systems. Following their design, Atelier could examine cases of project failure or missed deadlines on a one by one basis, similar to Upwork's dispute handling system. In such cases, Atelier may be able to replace mentors or interns who have dropped out or failed.

\subsection{Limitations}
Our research question was focused on how to enable learners to develop skill mastery. We attempted to measure ``mastery developed'' by comparing pre-task and post- tasks, but low statistical power made it difficult to make any conclusive statements --- a future goal is to be able to connect Atelier to longitudinal measures of growth and learning.

Another limitation of our deployment method was that, by virtue of repeating the same task across participants, the research team had to proxy for the requester. In a realistic deployment, the requester may have their own constraints, feedback, or evaluation criteria. However, in this study, different interns used various different approaches to complete the ECommerce project. Since the requester did not specify certain details, no approach was incorrect, and thus our expert evaluator had to make judgment calls. 

\subsection{Future Work}
Our main study examined the effects of Atelier in one domain: web development. To augment this, we also deployed Atelier with two pairs of interns and mentors to perform tasks outside the realm of web development, one creating a professional logo and the other writing a data mining crawler. Both successfully created their target product by taking advantage of Atelier. The quality of mentor feedback was similar to that given in the main study, and we found instances of sharing best practices.
\begin{quote}
Mentor: \textit{``indicate the color hex code for gray and green as the brand's colors. The client will ask for printing purposes, what typeface you used.''}
\end{quote}
However, we noticed that the logo design mentor limited their feedback to the technical and did not critique the creative aspects of the projects. Based on this observation, future work will design means to facilitate the creative process for jobs such as design and writing. For example, real-time brainstorming sessions would allow mentors to examine and improve the intern's thought process, and methods that explicitly support critique on creative work would ensure that mentors leave specific feedback on intern's work-in-progress.
\section{Conclusion}

In this paper, we have explored a model that uses existing crowdsourcing tasks as educational scaffolds for crowd experts to develop new skills and accumulate work experience through real world micro-internships. To demonstrate this idea, we presented Atelier, a platform that allows requesters to post jobs, connect with mentors, break down large tasks into achievable milestones, and mediate questions. Our evaluation confirms the potential benefits of Atelier's mentorship model: mentors provided instructional scaffolding by helping interns when they got stuck due to technical or conceptual issues, introducing best practices and new technologies, and sharing industry conventions. We envision that micro-internships will enable novice workers to take on challenging, new tasks, and, ultimately, to expand career opportunities in crowd work.

\section{Acknowledgements} 
This work was supported by a National Science Foundation CAREER award IIS-1351131, a Stanford Graduate Fellowship, and the Nakajima Foundation. 

\bibliographystyle{SIGCHI-Reference-Format}
\bibliography{references}

\balance{}

\end{document}